\documentclass{cimento}

\usepackage{graphicx}  

\title{Top Production at the Tevatron: The Antiproton Awakens}
\author{Kenneth Bloom for the CDF and D0 Collaborations}
\instlist{\inst Department of Physics and Astronomy, University of
Nebraska-Lincoln, Lincoln, Nebraska, USA}

\PACSes{
\PACSit{14.65.Ha}{top quarks}
\PACSit{13.88.+e}{polarization in interactions and scattering}
\PACSit{12.38.Qk}{experimental tests of quantum chromodynamics}
\PACSit{13.85.-t}{hadron-induced high- and super-high-energy interactions}
}

\begin{document}

\maketitle

\begin{abstract}
  A long time ago, at a laboratory far, far away, the Fermilab Tevatron collided
  protons and antiprotons at $\sqrt{s} = 1.96$~TeV.  The CDF and D0
  experiments each recorded datasets of about 10~fb$^{-1}$.  As such
  experiments may never be repeated, these are unique datasets that allow
  for unique measurements.  This presentation describes recent results from
  the two experiments on top-quark production rates, spin orientations, and
  production asymmetries, which are all probes of the $p\bar{p}$ initial
  state.
\end{abstract}

In 2001, the Fermilab Tevatron started colliding protons and antiprotons at
$\sqrt{s} = 1.96$~TeV.  When operations concluded in September 2011, two
experiments, CDF and D0, had each recorded datasets of approximately
10~fb$^{-1}$.  As it is possible that there will never again be a
$p\bar{p}$ collider, these are unique datasets that allow for unique
measurements.  In particular, the colliding partons were typically quarks
and antiquarks, in contrast to the gluon-gluon collisions that predominate
at the Large Hadron Collider (LHC).  Here we describe a number of recent
measurements that probe the production of top quarks at the Tevatron in
ways that cannot easily be explored at the LHC.  For the sake of brevity,
older measurements are only summarized while those that have yet to be
published or submitted for publication are described in greater detail.

Table~\ref{tab:xsec} gives the predicted production cross sections for
strongly-produced $t\bar{t}$ and electroweakly-produced single top quarks
at the Tevatron and at LHC.  While the cross sections are generally
much larger at the LHC, that is not the case for the $s$-channel single-top
production in the $tb$ mode, which is ``only'' a factor of five larger
than at the Tevatron.  Also, at the Tevatron the $q\bar{q}$ initial
state provides about 85\% of the total cross section for $t\bar{t}$
production, while it is only about 15\% (10\%) at the LHC in Run~1 (Run~2).
This allows the Tevatron to compete with the LHC in some areas, and
provides complementarity due to the $q\bar{q}$ initial state.

\begin{table}[t]
\begin{center}
\begin{tabular}{crrrr}  \hline
& $t\bar{t}$ & $tb$ & $tqb$ & $tW$ \\
Tevatron & 7.08 & 1.04 & 2.08 & 0.30\\
LHC Run 1 & 234 & 5.55 & 87.2 & 22.2\\
LHC Run 2 & 816 & 10.3 & 217 & 71.7\\\hline 
\end{tabular}
\caption{Cross sections, in picobarns, for the production of different
  final states including top quarks at the Tevatron ($p\bar{p}$, $\sqrt{s}
  = 1.96$~TeV) and the LHC Run~1 and Run~2 ($pp$, $\sqrt{s} = 8$~TeV and $\sqrt{s} = 13$~TeV), assuming a
  top-quark mass of 173~GeV~\cite{bib:kidonakis}.}
\label{tab:xsec}
\end{center}
\end{table}

\section{Production}
\begin{figure}[htb]
\centering
\includegraphics[height=2.0in]{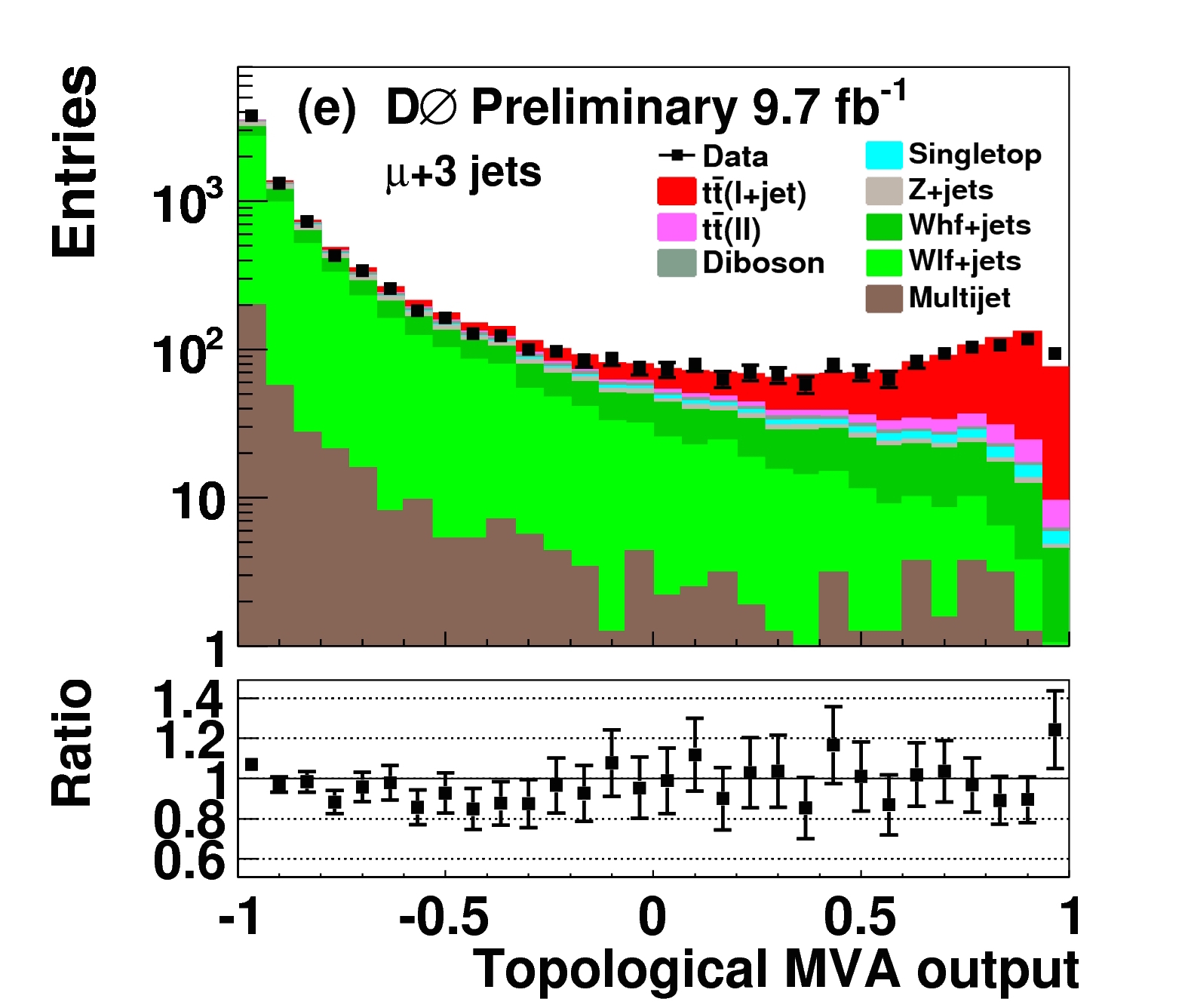}
\includegraphics[height=2.0in]{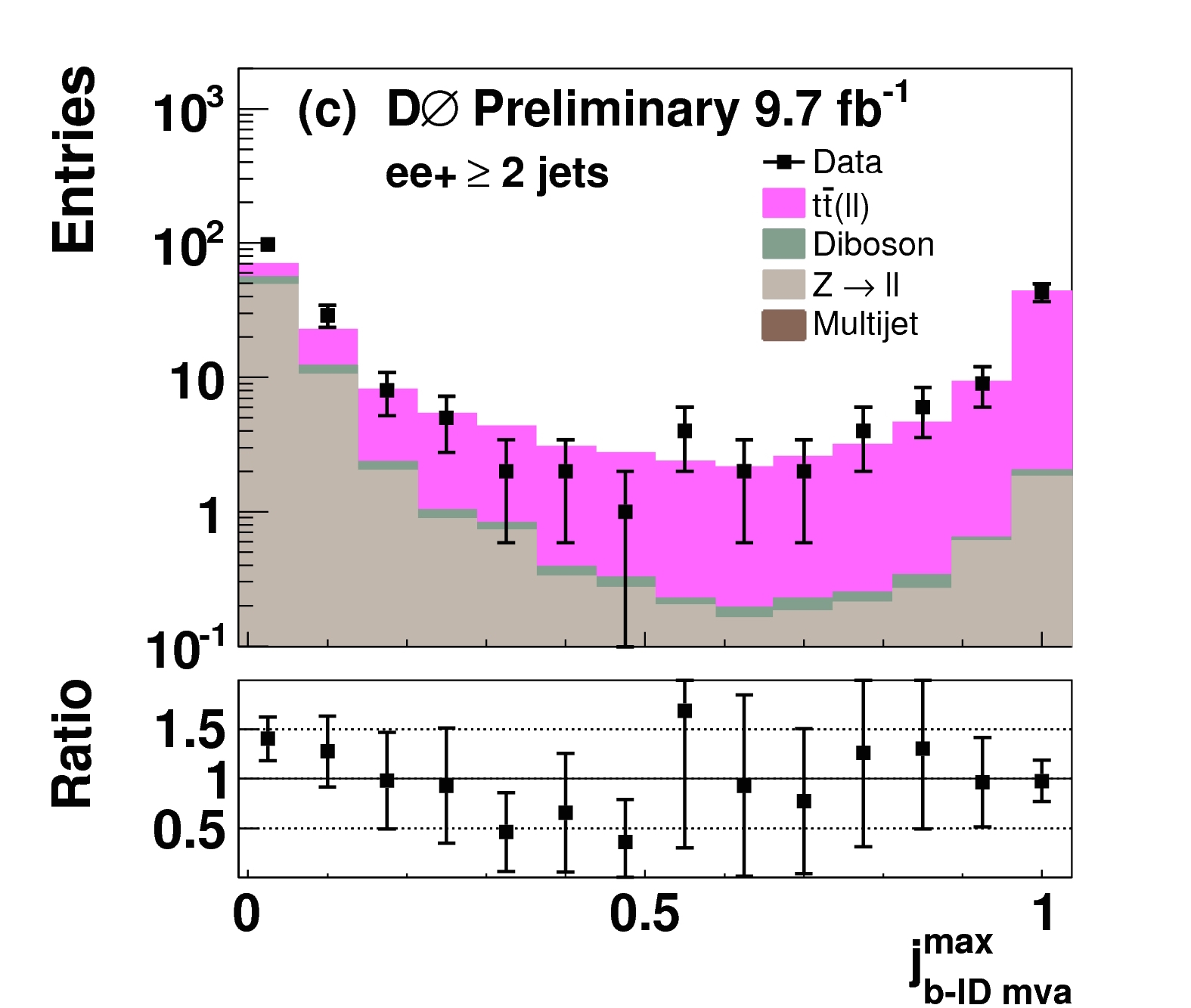}
\caption{Representative distributions from the D0 inclusive $t\bar{t}$ cross
  section measurement.  Left: Output of the boosted decision tree with
  gradients for the $\mu$ plus three jets channel.  Right: $b$-tag
  discrimination variable for the leading jet in the $ee$ plus at least two
  jets channel.  In both cases the colored histograms indicate the
  contributions from different physics proccesses.}
\label{fig:xsecvars}
\end{figure}

A preliminary measurement from D0, not yet published, gives a precise
measurement of the inclusive $t\bar{t}$ cross section that makes use of
both the dilepton and lepton-plus-jets channels~\cite{bib:ttxsec}. The
analysis makes heavy use of multivariate techniques, in which the numeric
values of many individual observables from an event are combined to form
one single quantity, and fits to distributions of those quantities from
each different final state are used to obtain the cross section.  The
lepton plus jets channel is broken into six subsamples based on lepton type
(electron or muon) and jet multiplicity (two, three or at least four jets).
Each subsample gets its own boosted decision tree with gradients using
about twenty kinematic variables, plus the output of a multivariate
algorithm used to identify $b$ jets.  The dilepton channel is simpler.  It
is broken into four subsamples ($e\mu$ plus one jet, $e\mu$ plus at least
two jets, $ee$ plus at least two jets and $\mu\mu$ plus at least two jets),
and the $b$-tag variable of the leading jet is the only one needed for the
fit.  Some representative distributions from the analysis are shown in
Figure~\ref{fig:xsecvars}.  The cross section is obtained from a
simultaneous log-likelihood fit template fit across all samples, using
systematic uncertainties as nuisance parameters.  The profiling of
systematic uncertainties reduces them by cross-calibration (for those that
are uncorrelated).  Careful attention is paid to correlations amongst
systematic uncertainties in the different subsamples.  The resulting cross
section is 7.73 $\pm$ 0.13 (stat) $\pm$ 0.55 (syst)~pb, where the leading
systematic uncertainties are from signal modeling, especially
hadronization.

As stated earlier, the Tevatron experiments have unique access to the
$s$-channel production of single top quarks; at the LHC, the backgrounds
(from $t\bar{t}$ production) are much more significant.  The rate for this
process is sufficiently small that results from the full datasets of both
experiments need to be combined to obtain a measurement with sufficient
statistical significance to be called an observation of the
process~\cite{bib:schannel}.  The cross section result,
$\sigma_{s \, \mathrm{channel}} = 1.29^{+0.26}_{-0.24}$~pb, has 6.3
standard deviations significance.  This measurement then allows separate
estimates of the $s$-channel and $t$-channel cross sections, without any
assumptions of the value of $\sigma_s/\sigma_t$.  The results are
consistent with the standard model predictions, with no indication of any
other contributing process.  The two cross section values then leads to a
measurement of $V_{tb}$ that makes no assumptions on the number of quark
generations, unitarity, or $\sigma_s/\sigma_t$ (but does assume standard
model top decays, a pure $V-A$ interaction, and CP conservation).  The
result is $|V_{tb}| = 1.02^{+0.06}_{-0.05}$, or $|V_{tb}| \geq 0.92$ at
95\% confidence level after applying a flat prior distribution for
$|V_{tb}| < 1$.

\section{Spin orientations}
Top quarks produced in the strong interaction are almost entirely
unpolarized, but for electroweak corrections at the 1\% level.  Thus, a
search for top polarization is a search for new physics.  The polarization
of the top quark can be measured in the top rest frame through angular
distributions of decay products $i$ with respect to a given axis $n$:
\begin{equation}
\frac{1}{\Gamma} \frac{d\Gamma}{d\cos\theta_{i,\hat{n}}} = 
\frac{1}{2} (1 + P_{\hat{n}}\kappa_i\cos\theta_{i,\hat{n}}),
\end{equation}
where $P_{\hat{n}}$ is the polarization strength and $\kappa_i$ is the
analyzing power of the decay product, which for leptons is nearly unity.
There are many axes $\hat{n}$ to choose from, such as the beam axis (the
direction of the proton), the helicity axis (the direction of the parent
top quark) and the transverse axis (the cross product of the other two
which is perpendicular to the production plane).  

\begin{figure}[htb]
\centering
\includegraphics[height=1.5in]{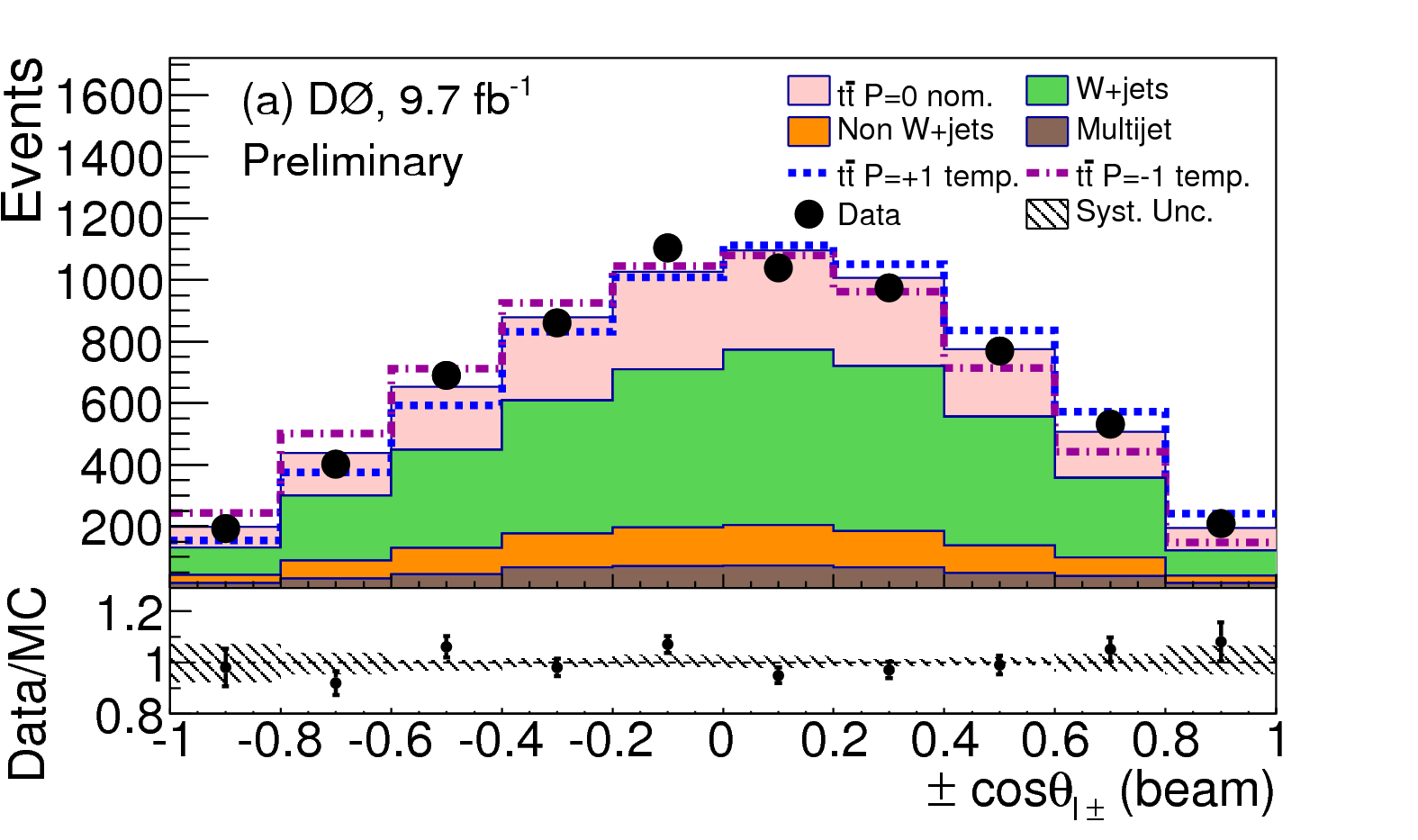}
\includegraphics[height=1.5in]{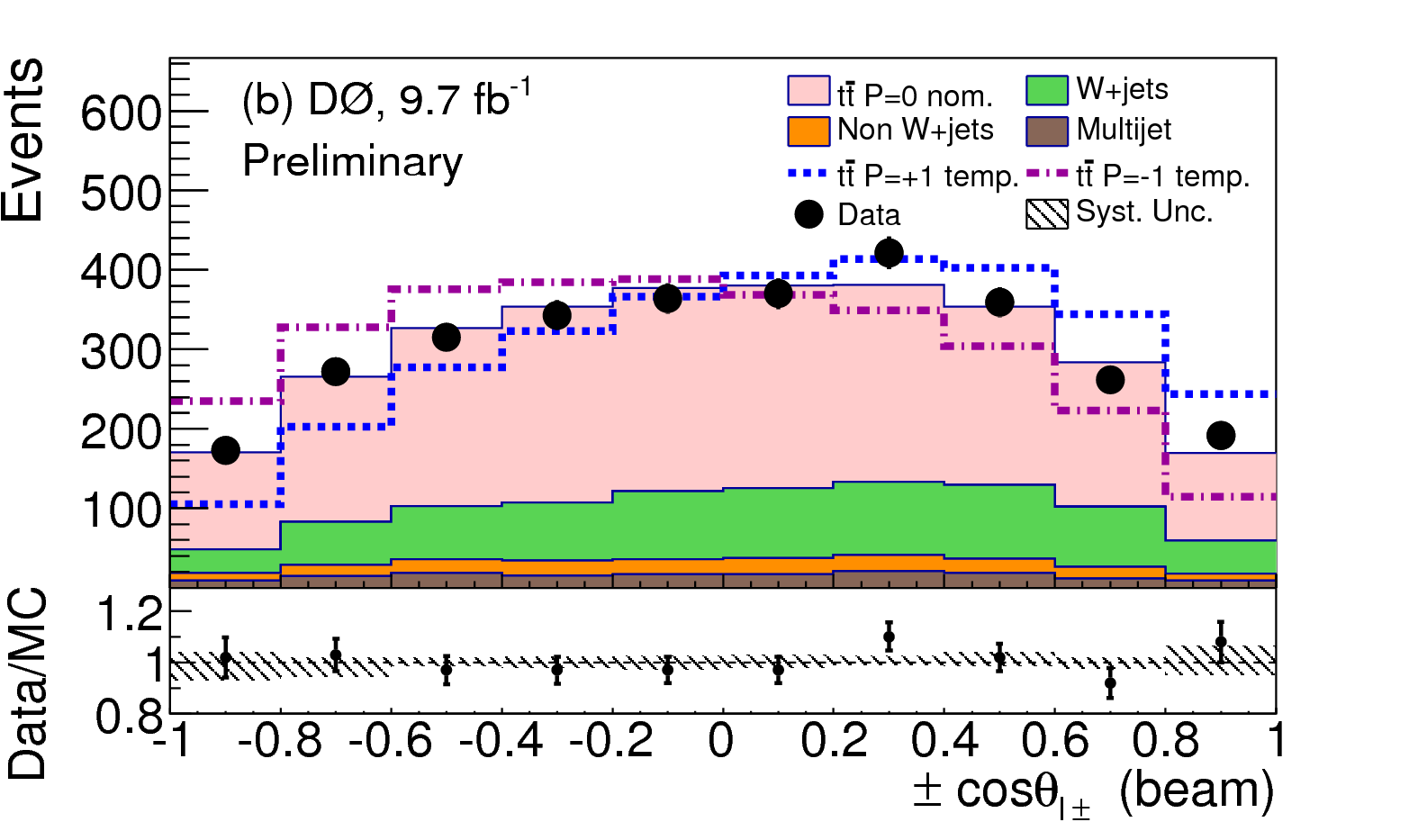}
\includegraphics[height=1.5in]{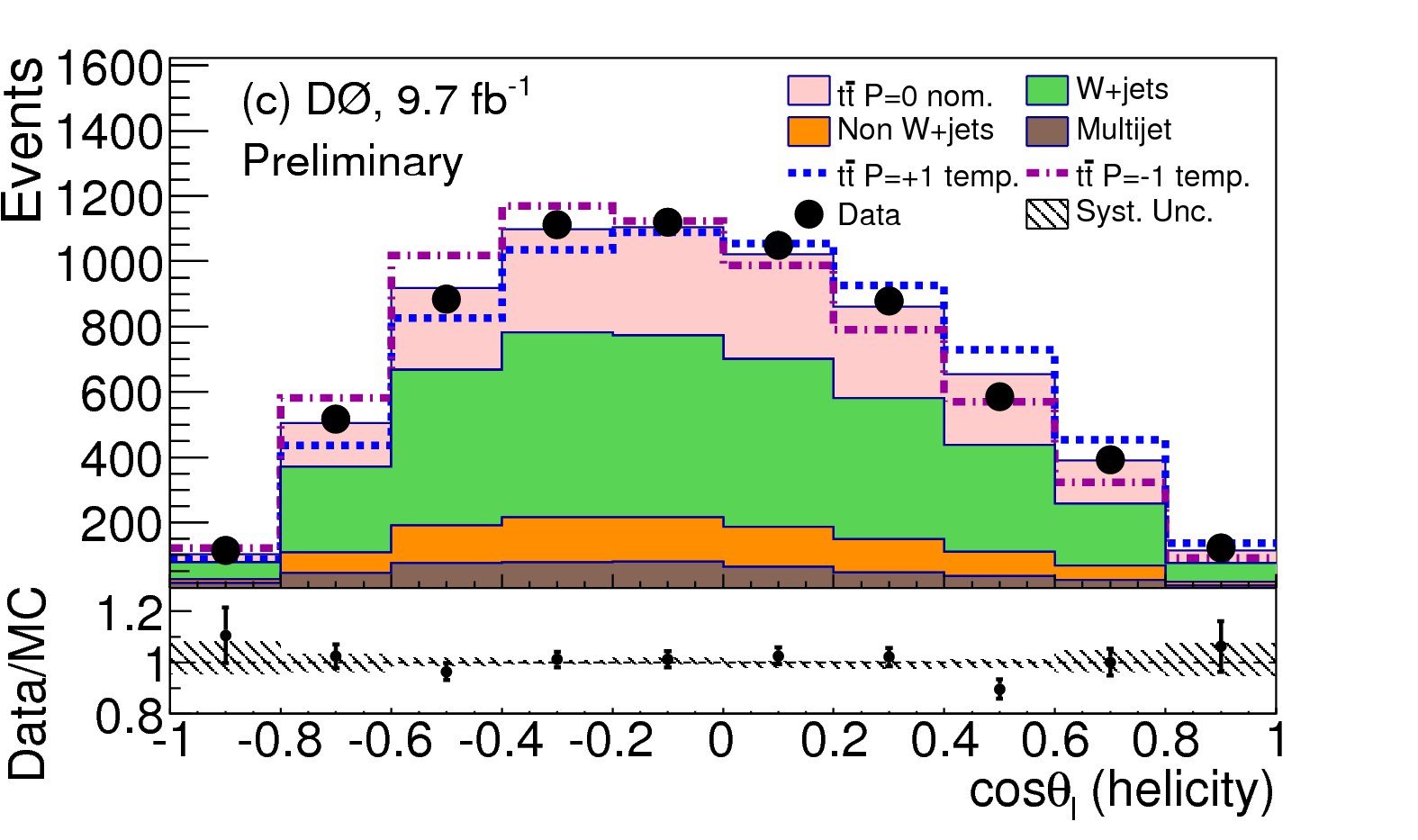}
\includegraphics[height=1.5in]{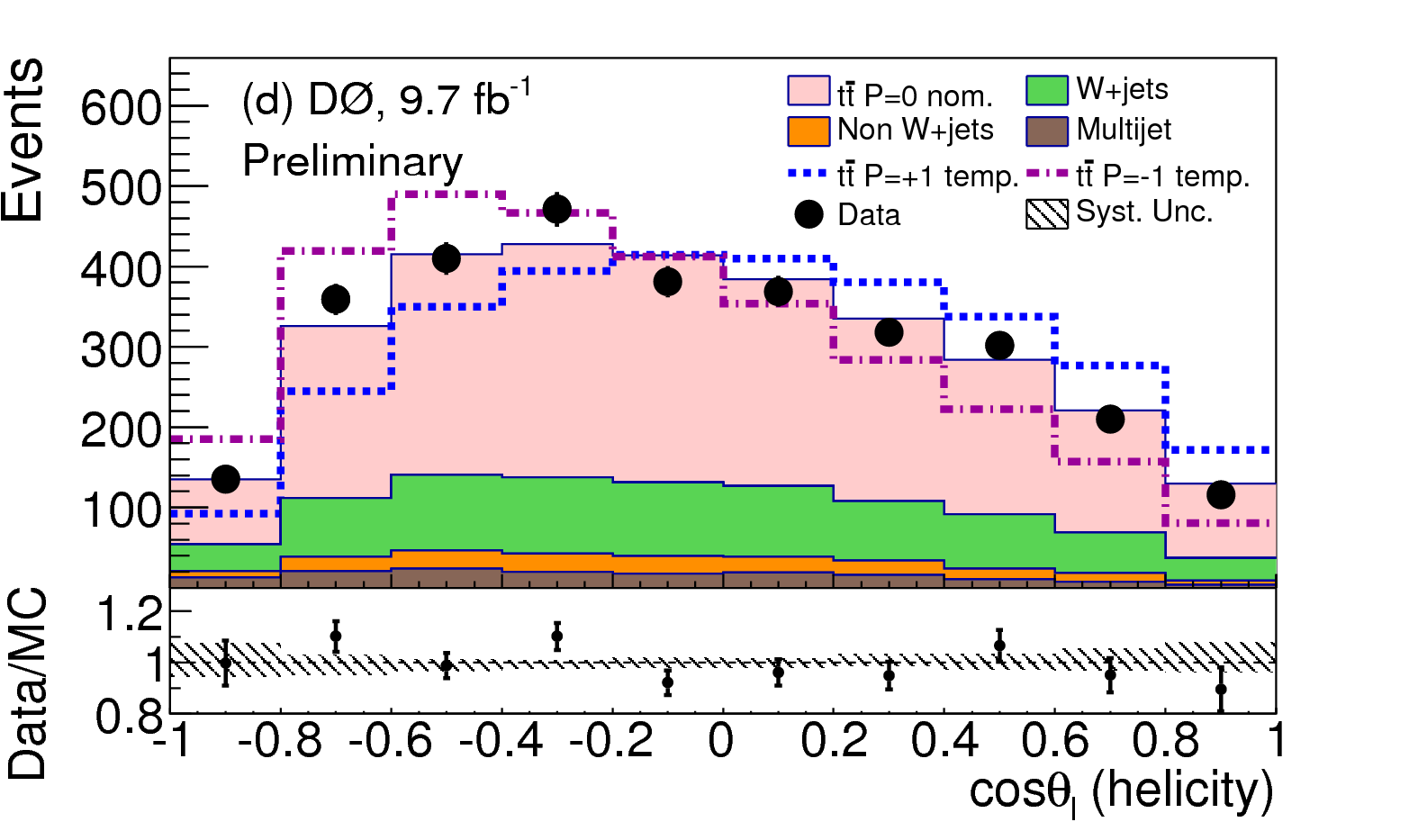}
\includegraphics[height=1.5in]{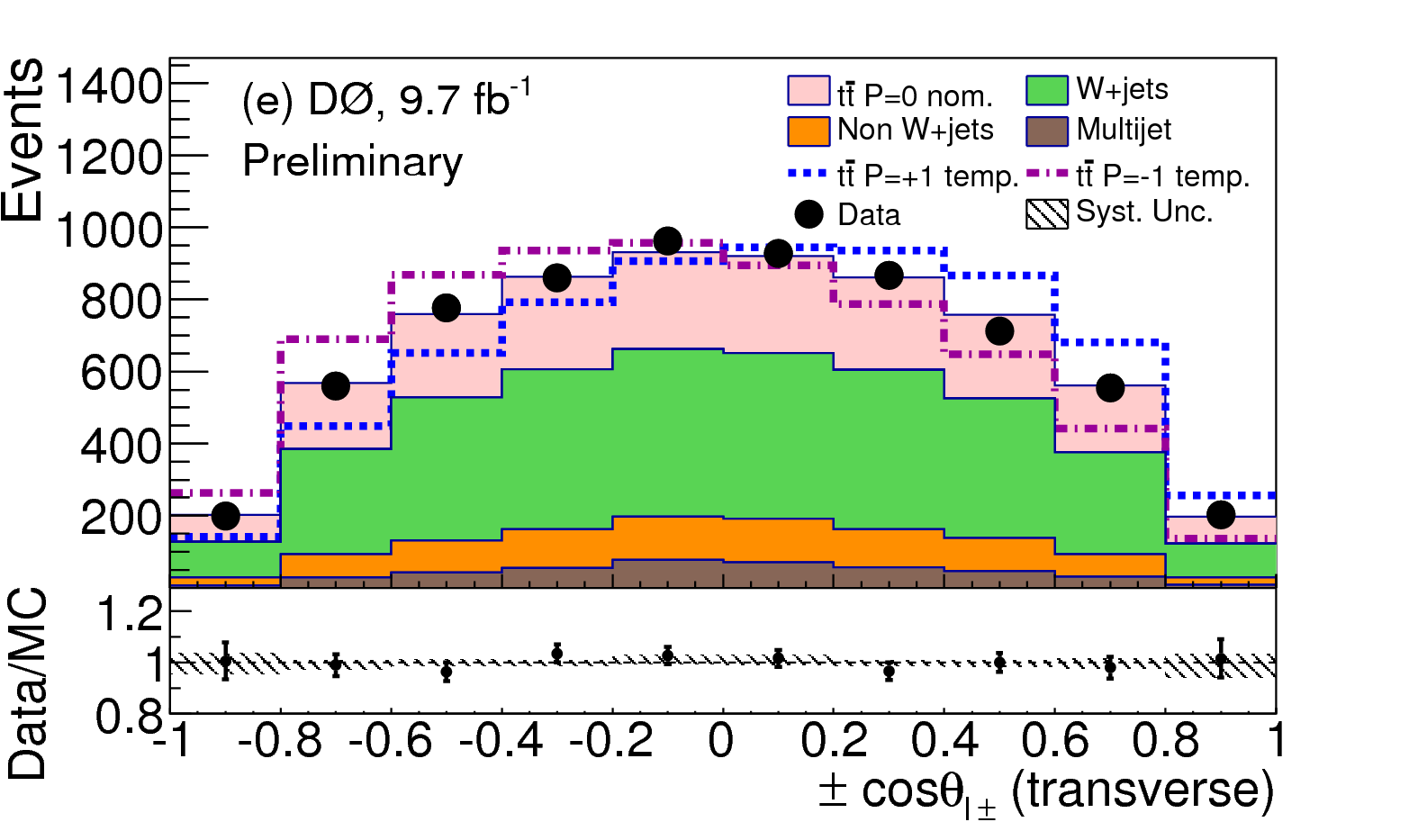}
\includegraphics[height=1.5in]{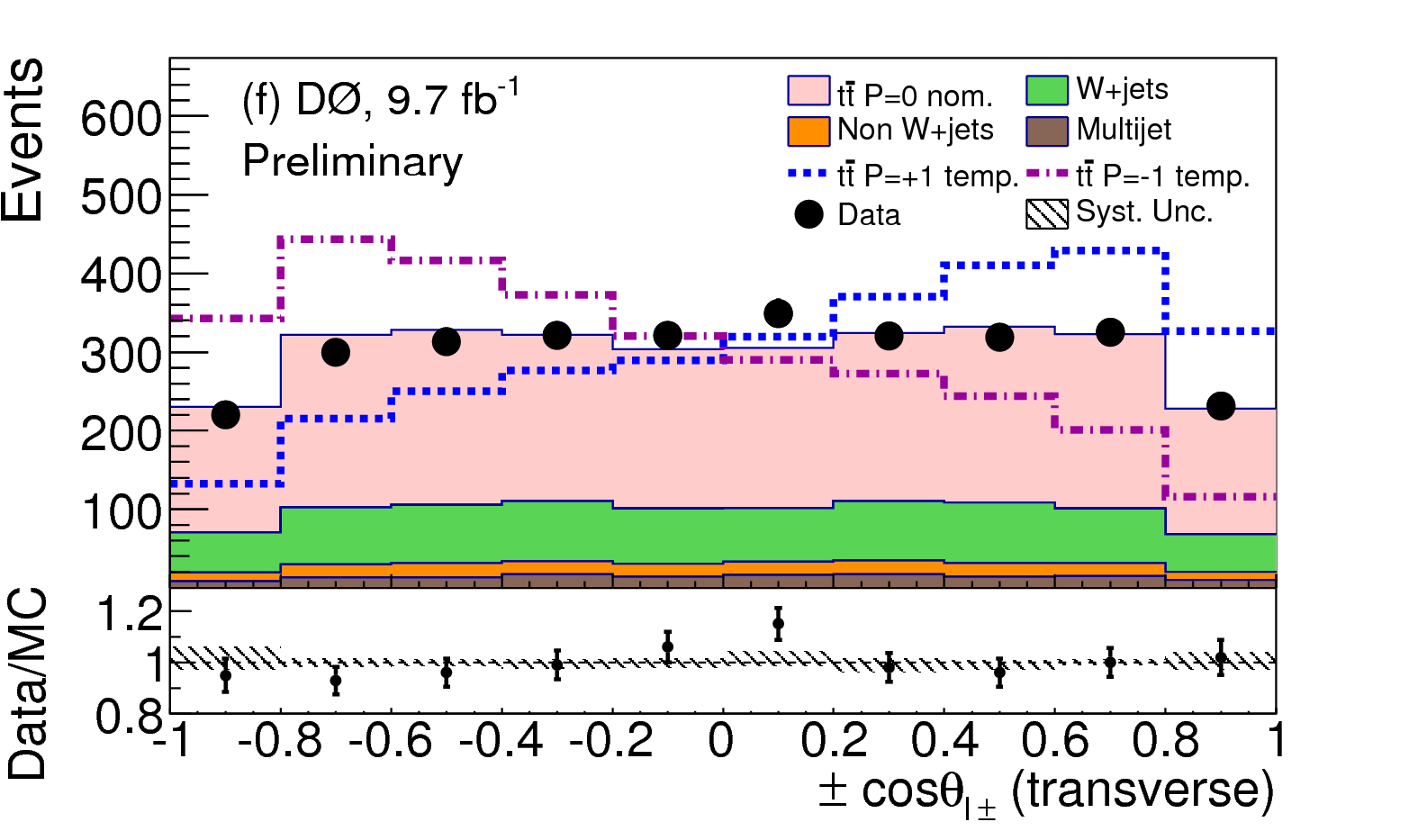}
\caption{The lepton plus jets $\cos\theta_{i,\hat{n}}$ distributions for
  data, expected backgrounds and signal templates for $P = −1,\, 0$ and
  $+1$. Panels (a), (c), and (e) represent selection with exactly three
  jets. (b), (d), and (f) represent selection with four or more jets. (a)
  and (b) show distributions in beam axis. (c) and (d) show distributions
  in helicity axis. (e) and (f) show distributions in transverse axis. The
  hashed area represents systematic uncertainty.}
\label{fig:pol}
\end{figure}

D0 has new measurements
of the top-quark polarization that makes use of the lepton plus three or
at least four jets samples, in which a kinematic reconstruction is done is performed
to obtain the lepton angles $\cos\theta_{i,\hat{n}}$~\cite{bib:D0pol}.  The
inclusion of the three-jet sample increases the statistical power of the
measurement but requires the use of a kinematic fitter developed for the
$A_{FB}$ measurement described in Section~\ref{sec:afb}.  Templates in
$\cos\theta_{i,\hat{n}}$ for $P = +1$ and $P = -1$ are developed, and a fit
of the angular distributions to the two templates and a background shape is
performed.  The relative normalizations of the templates allow for the
extraction of $P_{\hat{n}}$.  The $\cos\theta_{i,\hat{n}}$ distributions
and fit results are shown in Figure~\ref{fig:pol} and Table~\ref{tab:pol}.
The measurements are consistent with both the standard model prediction and
zero, and the measurement of the transverse polarization is the first ever.
A measurement of the polarization in dilepton events is described in
Section~\ref{sec:afb}.

\begin{table}[t]
\begin{center}
\begin{tabular}{ccc}  \hline
Axis & Measured polarization $P_{\hat{n}}$ & SM prediction\\\hline
Beam & $+0.070 \pm 0.055$ & -0.002\\
Helicity & $-0.102 \pm 0.060$ & -0.004\\
Transverse & $+0.040 \pm 0.034$ & +0.011\\\hline
\end{tabular}
\caption{Measured top quark polarization in beam, helicity, and transverse
  spin quantization bases along with the standard model (SM)
  predictions~\cite{bib:poltheory}}
\label{tab:pol}
\end{center}
\end{table}

While $t\bar{t}$ pairs are not produced polarized, their spins are
correlated.  The measurement of this correlation is unique to the
$t\bar{t}$ system, as the top lifetime is a thousand times shorter than the
spin decorrelation time.  The amount of correlation depends on the initial
state, $q\bar{q}$ or $gg$.  D0 has measured the spin correlation using the
observable
\begin{equation}
O_{\rm off} = 
\frac{\sigma(\uparrow\uparrow) + \sigma(\downarrow\downarrow) 
- \sigma(\uparrow\downarrow) - \sigma(\downarrow\uparrow)}
{\sigma(\uparrow\uparrow) + \sigma(\downarrow\downarrow) +
  \sigma(\uparrow\downarrow) + \sigma(\downarrow \uparrow)}
\end{equation}
using the ``off-diagonal'' spin quantization basis~\cite{bib:offdiagonal}
where the correlation is maximized~\cite{bib:d0spincorr}.  Both dilepton
and lepton+jets events are reconstructed with the matrix element method to
create a discriminant distribution that reflects the relative probability
for the SM or null spin correlation hypotheses.  The resulting measurement
is $O_{\rm off} = 0.89 \pm 0.16\, {\rm (stat)} \pm 0.15\, {\rm (syst)}$,
where the systematic uncertainties are dominated by signal modeling issues.
The measured value is in agreement with the SM value of
0.80~\cite{bib:SMspincorr}, and is 4.2 standard deviations away from zero,
giving evidence for spin correlation.  In addition, as the $q\bar{q}$ and
$gg$ initial states lead to different correlation strengths, the fraction
from $t\bar{t}$ from each initial state at next to leading order can be
extracted.  The result is
$f_{gg} = 0.08 \pm 0.12\, {\rm (stat)} \pm 0.11\, {\rm (syst)}$, consistent
with the SM value of $0.135$~\cite{bib:SMspincorr}.

\section{Production asymmetries}
\label{sec:afb}
Due to interference terms that arise at next to leading order in QCD,
$t\bar{t}$ pairs produced from $q\bar{q}$ interactions have a
forward-backward asymmetry in the direction of the resulting quarks; the
$t$ tends to follow the direction of the $q$ and the $\bar{t}$ the
direction of the $\bar{q}$.  This asymmetry, defined as
\begin{equation}
A_{FB} = \frac{N(\Delta y > 0) - N(\Delta y < 0)}
{N(\Delta y > 0) + N(\Delta y < 0)},
\end{equation}
where $\Delta y = y_t - y_{\bar{t}}$, the rapidity difference between the
top quark and antiquark, is predicted to be about
10\%~\cite{bib:xsecvsmasscalc}.  The Tevatron has unique access to this
quantity as most of the $t\bar{t}$ pairs are produced in $q\bar{q}$
interactions, which is not the case at the LHC.  

The $t\bar{t}$ forward-backward asymmetry has been a topic of great
interest for some years, as an anomalously large result could be an
indicator of new physics, and some early measurements of this quantity
using the amount of Tevatron data that was available at the time were in
fact quite large (and the predicted values had been smaller).  That spawned
an effort to probe the asymmetry through a number of measurements that are
briefly described here.

Reconstructing the top direction for $A_{FB}$ is complicated, as it requires a
kinematic reconstruction and then an unfolding because the experimental
resolution is poor.  Another approach to the problem is to measure the
forward-backward asymmetry of the decay lepton.  While the SM prediction
for the asymmetry is only 4\%, the measurement is relatively simple because
of the resolution on the lepton direction.  CDF has measured
$(9.0^{+2.8}_{-2.6})$\%~\cite{bib:CDFal} and D0 has measured
$(4.2 \pm 2.4)$\%~\cite{bib:D0al} for this quantity.  The CDF result is
slightly above the expected value, with an observed dependence on the
lepton rapidity.

Another approach to the problem is to look in a different system.
New physics affecting $A_{FB}$ should affect $b\bar{b}$ production too.
Most $b\bar{b}$ production is from the $gg$ initial state, but $q\bar{q}$
production is enhanced for high-mass pairs.  CDF has made two measurements
of the $b\bar{b}$ asymmetry.  One focuses on high-mass pairs, identifying
$b$ jets with secondary vertices and assigning flavor with the difference
in measured jet charges between the two jets.  Effects that dilute the
asymmetry such as mixing, secondary decays, charge misidentification and
non-$b$ backgrounds are accounted for and the result is unfolded to the
particle level.  The result is consistent with the SM and is able to
exclude some axigluon models~\cite{bib:CDFbbhigh}.  A more recent 
search using lower-mass $b\bar{b}$ pairs makes use of soft-muon tagging to
identify the $b$ jets, and is also consistent with SM
expectations~\cite{bib:CDFlow}.

But the most fundamental information is obtained from the $A_{FB}$
measurements themselves, in which the top quark directions are
reconstructed.  Both Tevatron experiments have well-established
measurements in the lepton plus jets samples.  CDF measures $A_{FB} = (16.4
\pm 4.5)\%$, somewhat higher than the SM prediction~\cite{bib:CDFAfbljets}.
D0's measurement uses more the phase space by exploiting the three-jet
sample, along with a new top reconstruction method and two-dimensional
unfolding.  The result, $A_{FB} = (10.6 \pm 3.0)\%$, is more consistent
with the SM prediction~\cite{bib:D0Afbljets}.  Both experiments examine the
dependence of $A_{FB}$ on $m_{t\bar{t}}$ and $|\Delta y|$ of the quarks,
and find that it is greater than predicted.

$A_{FB}$ measurements in the dilepton sample are more challenging because
of the two neutrinos in the final state, and took longer to complete.  The
recently published D0 analysis~\cite{bib:D0Afbdil} measures the production
asymmetry simultaneously with the polarization of the top quark with
respect to the beam axis, using a novel application of the matrix-element
technique.  A full reconstruction of the event kinematics is performed in a
probabilistic fashion, and then a likelihood per event for the most
probable kinematic value is made for both the asymmetry and the lepton
decay angle with respect to the beam axis in the $t\bar{t}$ rest frame.
After an appropriate calibration of the method, the relevant quantities can
be extracted from the distributions of these quantities. The systematic
uncertainties are dominated by those involved in modeling the $t\bar{t}$
signal, in particular hadronization and showering, and also the calibration
of the method.

Without constraining either the asymmetry or the polarization, the results
are
\begin{eqnarray}
A_{FB} &=& (15.0 \pm 6.4 \pm 4.9)\%\\
\kappa P &=& (7.2 \pm 10.5 \pm 4.2)\%,
\end{eqnarray}
where the first uncertainty is statistical and the second is systematic,
and $\kappa \simeq 1.0$ is the spin analyzing power of
the lepton.  If one of the quantities is constrained to its standard-model
value, the result for the other quantity is
\begin{eqnarray}
A_{FB} &=& (17.5 \pm 5.6 \pm 3.1)\%\\
\kappa P &=& (11.3 \pm 9.1 \pm 1.9)\%.
\end{eqnarray}
The latter result for $A_{FB}$ is combined with that from the lepton plus
jets measurement to obtain the final D0 measurement of this quantity,
$A_{FB} = (11.8 \pm 2.5 \pm 1.3)\%$.

\begin{figure}[htb]
\centering
\includegraphics[height=1.5in]{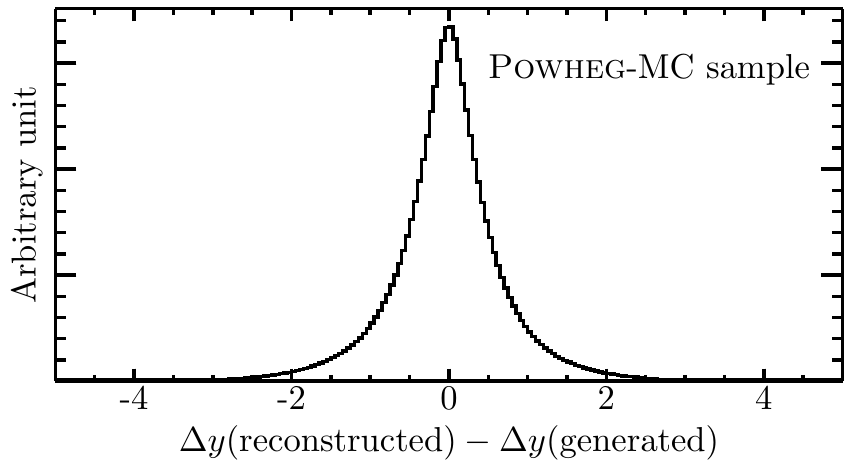}
\includegraphics[height=1.5in]{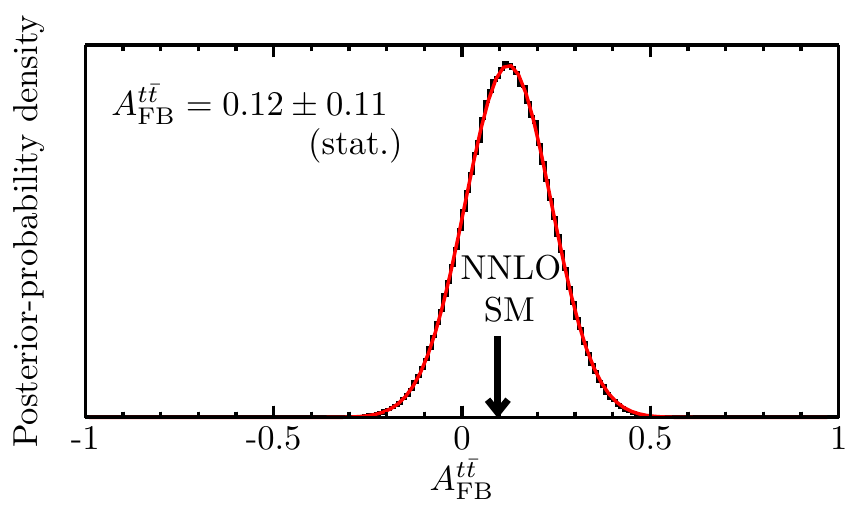}
\caption{Left: Distribution of $|\Delta y$(reconstructed)-$|\Delta
y|$(generated) in the CDF dilepton $A_{FB}$ measurement.  Right: The
resulting posterior probability for $A_{FB}$.}
\label{fig:CDFAfbdil}
\end{figure}

A CDF measurement of $A_{FB}$ in the dilepton final state has recently been
submitted for publication~\cite{bib:CDFAfbdil}.  It is carried out in the
same spirit as the D0 measurement.  A likelihood-based algorithm is used to
reconstruct the momenta of the two neutrinos, and thus the top momenta, in
each event from the observed kinematics.  Rather than a single solution, a
likelihood is formed as a function of the kinematic variables, and both
possible lepton-jet pairings are included.  A likelihood-based scheme is
used to unfold the results back to the parton level.  The event selection
is optimized to avoid poorly-reconstructed events, which keeps the
migration matrix fairly diagonal.  Figure~\ref{fig:CDFAfbdil} shows the
expected resolution on the measurement of $\Delta y$, and the posterior
probability density obtained from the event sample.  The resulting value of
$A_{FB}$ is $(12 \pm 11 \pm 7)$\%, where the first uncertainty is
statistical and the second is systematic.  While there is some sensitivity
to the $|\Delta y|$ dependence of the result, no significant dependence is
observed.  The result is then combined with the CDF lepton plus jets result
to obtain $A_{FB} = (16.0 \pm 4.5)$\%, consistent with the SM expectation.

\begin{figure}[htb]
\centering
\includegraphics[height=3in]{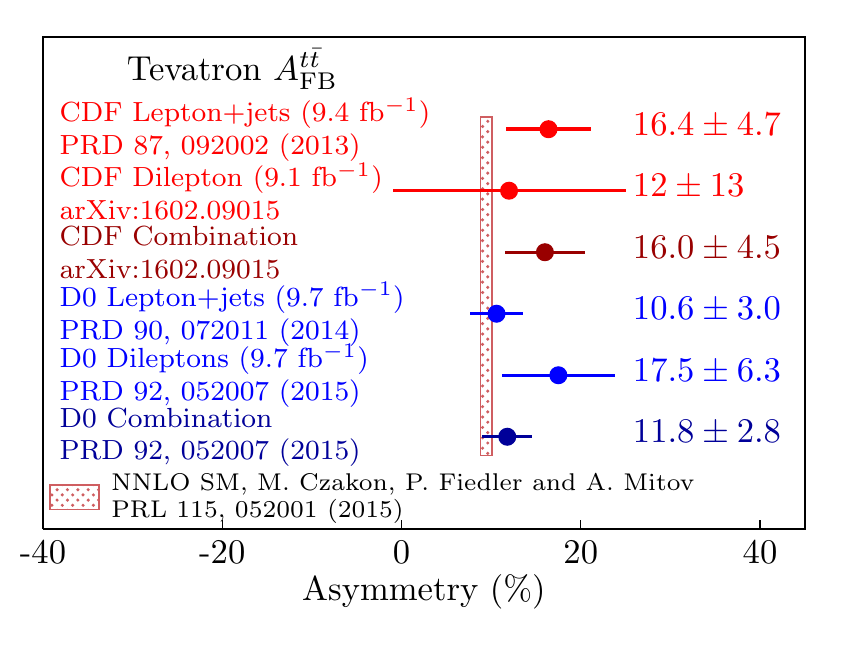}
\caption{Summary of $A_{FB}$ measurements from CDF and D0.}
\label{fig:Afbsummary}
\end{figure}

A summary the final Tevatron measurements of $A_{FB}$, using the full
datasets, is given in Figure~\ref{fig:Afbsummary}.  The agreement between
the results from CDF and D0 is reasonable, as is the agreement with the
predictions from the standard model.

\section{Conclusions}
Even with the onslaught of data from the LHC, top physics at the Tevatron
has remained interesting.  The complementarity of the proton-antiproton
initial state of the Tevatron has provided unique opporunities.  The
production asymmetry cannot be explored as well at the LHC, and the
$s$-channel single-top production is much more difficult to study there
too.  In addition, CDF and D0 are very mature experiments, with
well-understood datasets and well-modeled detectors.  This allows for
significant creativity in data analyses that have yielded sophisticated
measurements.  The $A_{FB}$ measurements in particular drove a spectacular
effort to fully exploit the capabilities of the two experiments.  Arguably
the LHC has much to learn from the Tevatron experience.  The last few
Tevatron top production measurements should soon be available, bringing
this epic adventure to a conclusion.

\acknowledgments 
Thousands of physicists worked on the Tevatron and on the
CDF and D0 experiments over decades, and the results presented are due to
their efforts.  I particularly thank the current top-physics group
conveners of the experiments for their input and feedback on their
presentation.  I also thank the conference organizers for the opportunity
to enjoy La Thuile.  May the quarks be with you!

\end{document}